\documentstyle[12pt]{article}
\newcommand{\h}{\hspace{.5cm}}
\newenvironment{destaque}{\begin{quotation}\small\em}{\end{quotation}}
\date{}
\title{Demonstration of how the zeta function method for effective potential removes the divergences}
\author{{\bf Jos\'e Alexandre Nogueira}\\
{\it Departamento de F\'{\i}sica, Centro de Ci\^encias Exatas,}\\
{\it Universidade Federal do Espirito Santo,}\\
{\it 29.060-900 - Vit\'oria-ES - Brasil,}\\
{\it E-mail: } nogueira@cce.ufes.br\\
{\bf Adolfo Maia Jr.}\\
{\it Instituto de Matem\'atica, Universidade Estadual de Campinas,}\\
{\it 13.081-970 - Campinas-SP - Brasil}\\
{\it E-mail: } maia@ime.unicamp.br}
\begin{document}
\maketitle

{\bf Abstract}
\begin{destaque}
The calculation of the minimum of the effective potential using the zeta function method is extremely advantagous, because the zeta function is regular at $s=0$ and we gain immediately a finite result for the effective potential without the necessity of subtratction of any pole or the addition of infinite counter-terms. The purpose of this paper is to explicitly point out how the cancellation of the divergences occurs and that the zeta function method implicitly uses the same procedure used by Bollini-Giambiagi and Salam-Strathdee in order to gain finite part of functions with a simple pole.
\end{destaque}

\section{Introduction}

\h High-energy physics was the first to perceive the necessity for a relativistic quantum field theory, and later other speciality physics found a powerful tool in it. Undoubtedly the quantum field theory obtained its first sucess in the quantum electrodynamics (QED) and its best sucess in the standard model. Quantum field theory is fundamentaly of perturbative aspect. So, the quantities with the greatest physics interest, the Green functions, are constructed by means of perturbative series. However, in all its perturbative aspects it has heavy divergence problems. The treatment of these infinites continues to be one of the most challenging problems in quantum field theory. The mathematical nature of the problem is clear. Divergences occur in perturbative computations because two distribuitions can not be multiplied at the same point. Several methods have been proposed in order to eliminate this problem. However, only has it been possible to eliminate these infinites consistently and in a physically meaningful manner by absorbing them into the bare parameters of the theory.

A quantity of considerable physics importance is the vacuum energy density which is associated with interesting physics effects, such as, Lamb Shift and Casimir Effect, which occur because of vacuum fluctuation. Nevertheless, there are several variations on the concept of vacuum energy in commom circulation, among them the minimum of the effective potential obtained from the approach of functional methods from quantum field theory is largely used. Effective potential principal application is associated with spontaneous symmetry breaking. It is obtained from a nonperturbative method as a series in loop ($\hbar$). Since in the classical limit, which is the tree approximation, the effective potential becomes the same as the classical potential, therefore it is the classical potential plus the quantum corrections. It also suffers from the same divergences problems.

The usual procedure in order to deal with those divergences has been to employ a regularization method (Dimensional, Cut-off, etc) so as to isolate the divergences and to become the finite theory making use of a regulator and afterwards using a renormalization prescription, subtraction of the poles or addition of counter-terms, to eliminate the isolated divergences and to restore the original theory with the elimination of the regulator. Since the substraction of the poles or addition of infinite counter-terms, although well-founded in flat spaces, become dubious in curved spaces, the calculation of the minimum of the effective potential using the zeta function method is advantagous. Since the zeta function is regular at $s=0$ and we get a finite result for the effective potential without the apparent necessity of subtratction of any pole or the addition of infinite counter-terms. However, it is obvious that implicitly in the zeta function method there must takes place cancelling of the divergences.

The paper is organized as follows. In Section 2 we explicitly show the cancellation of the divergences in the zeta function method. In Section 3 we show that the zeta function method is equivalent to a well-known procedure to obtain finite part of functions with a simple pole (in this case is one related to the zero-point energy) used  by Bollini-Giambigi and Salam-Strathdee. This is because the zeta function method is an analytic regularization procedure. In Sections 4 we point out our conclusions.

\section{Cancellation of the divergences}

\h In the approach of the functional methods from quantum field theory, the vacuum energy density can be found by computing the minimum of the effective potential [1 - 7]. The energy density found of that manner is a loop expansion (or equivalently in powers of $\hbar$), that is, its classical amount plus quantum correction.

Let $\phi(x)$ be a single real scalar field in a  Minkowski
space-time, subject to the potential $V(\phi)$. The minimum effective potential to the first order in the loop expansion (or equivalently
in powers of $\hbar$) is given by 
\begin{eqnarray}
V_{ef}(\bar{\phi}) = V_{cl}(\bar{\phi}) + \frac{1}{2}\frac{\hbar}{\Omega}\ln\det\biggl[
\frac{\delta^{2}S[\bar{\phi}]}{\delta\phi(x)\delta\phi(y)}\biggr] =
V_{cl}(\bar{\phi}) + V_{ef}^{(1)}(\bar{\phi}),
\end{eqnarray}
where $\bar{\phi}=<\phi>$ is the vacuum expectation value, $S[\phi]$ is the
classical action, $\Omega=VT$ is the volume of the background
space-time manifold and in the classical potential $V_{cl}(\phi)$ is
included mass and self-interactions terms.

Making usual analytic continuation to the Euclidian space-time [2, 4], the classical action can be written as
\begin{eqnarray}
S[\phi]= \int d^{4}x\biggl[\frac{1}{2}\partial_{\mu}\phi\partial_{\mu}\phi
+ V_{cl}(\phi)\biggr],
\end{eqnarray}
where an Euclidian summation convention is understood for repeated
indexes. From eq.(2) we get the matrix ${\it m}(x,y)$ of the quadratic variation of the action $S[\phi]$
\begin{eqnarray}
{\it m}(x,y) \equiv \frac{\delta^{2}S[\phi_{c}]}{\delta\phi(x)
\delta\phi(y)} = \delta^{4}(x-y)[-\delta^{\mu\nu}\partial_{\mu}
\partial_{\nu} + V_{cl}^{\prime\prime}(\phi_{c})].
\end{eqnarray}
The classical field $\phi_{cl}(x) = <\phi(x)>_{J}$ is the vacuum expectation in the presence of an external source $J(x)$. When $J(x) \rightarrow 0$, $\phi_{cl}(x)$ becomes a constant, $\phi_{c}$.

Now, {\it m} is a real, elliptical and self-adjoint operator (because of the Euclidean analytic continuation) and for these kind of operators we can 
define the so-called generalized zeta function. Let $\{\lambda_{i}\}$ the eigenvalues of the operator $m(x,y)$. The generalized zeta function associated to ${\it M}(x,y)$ $(m \rightarrow {\it M} = \frac{m}{2\pi\mu^{2}})$ is defined by 
\begin{eqnarray}
\zeta_{\it M}(s) = \frac{1}{2}\frac{\hbar}{\Omega}\sum_{i}\biggl(\frac{\lambda_{i}}{2\pi\mu^{2}}\biggr)^{-s},
\end{eqnarray}
where we have introduced a unknown scale parameter $\mu$, with the dimensions of (length)$^{-1}$ or mass in order to keep the zeta function dimensionless for all s. The introduction of the scale parameter $\mu$, can be best understood when we observe that a hidden splitting of the divergent integral there is in the proceeding of zeta function regularization, that is, a separation of the divergent and finite parts of the
$V_{ef}(\phi_{c})$ (pag. 208 [4] and pag. 88 [5]. It is well-known the relation [2] 
\begin{eqnarray}
\ln\det{\it M} = - \frac{d\zeta_{\it M}(0)}{ds}.
\end{eqnarray}
Now, effective potential to the first order in the loop expansion can be written
\begin{eqnarray}
V_{ef}^{(1)}(\phi_{c}) = - \frac{d\zeta_{\it M}(0)}{ds}.
\end{eqnarray}

The evaluation of the effective potential using of the eq.(6) yields a finite result without the necessity of subtraction of any pole or the addition of infinite counter-terms. This is because the generalized zeta function as definided in the eq.(4) is regular at $s=0$ [2, 8]. Evidently the necessity to fit the theory parameters to observed results lead us to impose renormalization conditions, such as
\begin{equation}
\left. \frac{d^{2}V_{ef}}{d\phi _{c}^{2}}\right| _{\left\langle
\phi \right\rangle }=m_{R}^{2},
\end{equation}
\begin{equation}
\left. \frac{d^{4}V_{ef}}{d\phi _{c}^{4}}\right| _{\left\langle
\phi \right\rangle }=\lambda _{R},
\end{equation}
where $m_{R}$ is the renormalized mass, $\lambda_{R}$ is the renormalized coupling constant and $<\phi>$ is the minimum point of the effective potential (subtract point) [6, 9].

In the case of theories of null mass, the subtraction point for the
renormalization condition (8) cannot be taken at $\left\langle \phi
\right\rangle =0$ due to the logarithmic singularity. Even so in that
 case there is no intrinsic mass scale; therefore all the 
renormalization points are equivalent and the condition (8) is 
replaced by 
\begin{equation}
\left. \frac{d^{4}V_{ef}}{d\phi _{c}^{4}}\right| _{\left\langle
\phi \right\rangle = M}
=\lambda _{R},
\end{equation}
where $M$ is a arbitrary floating renormalization point [4].

Alternatively we use the relation
\begin{eqnarray}
\ln\det[m(x,y)] = tr\ln[m(x,y)],
\end{eqnarray}
and get [4]
\begin{eqnarray}
V^{(1)}_{ef}(\phi_{c}) = \frac{\hbar}{2}\int\frac{d^{4}k}{(2\pi)^{4}}\ln\biggl[k^{2}_{E} + V^{\prime\prime}(\phi_{c})\biggr],
\end{eqnarray}
where $k_{E}$ is the quadri-moment and the lower suffixes E denotes Euclidian space-time. As can be seen the integral of the eq.(11) is clearly divergent and so we need a regularization procedure in order to isolate the divergences. Of this we conclude that the evaluation of the effective potential using the zeta function, eq.(6), must hide the cancellation of the divergences in some manner.

In order to explicitly point out how the cancellation of the divergences occurs, we write the eigenvalues of the operator $m(x,y)$ as
\begin{eqnarray}
\lambda_{i,\omega} = \omega^{2} + h_{i}^{2},
\end{eqnarray}
where $h_{i}$ are eigenvalues of the Hamiltonian operator $H$ and $\omega$ is a continuous parameter labeling the temporal part of the eigenvalues of the operator $m(x,y)$.

Generalized zeta function associated to the operator ${\it M}(x,y)$, definited by eq.(4), can be written, using the eq.(12), as
\begin{eqnarray}
\zeta_{M}(s) = \frac{1}{2}\frac{\hbar}{\Omega}\int_{-\infty}^{\infty} \frac{d\omega}{2\pi}\sum_{i}\biggl[\frac{\omega^{2}}{2\pi\mu^{2}} + \frac{h_{i}^{2}}{2\pi\mu^{2}}\biggr]^{-s}T.
\end{eqnarray}
Using the relation [10]
\begin{eqnarray}
\int_{-\infty}^{\infty}\biggl(k^{2} + A^{2}\biggr)^{-s}d^{m}k =
\frac{\pi^{\frac{m}{2}}\Gamma(s-m/2)}{\Gamma(s)}\biggl(A^{2}
\biggr)^{\frac{m}{2}-s},
\end{eqnarray}
we can perform the above integral in $d\omega$ and get
\begin{eqnarray}
\zeta_{M}(s) = \frac{1}{2\sqrt{\pi}}\frac{\Gamma[s-1/2]}{\Gamma[s]}\frac{1}{2}\frac{\hbar}{\Omega}\zeta_{H}(s-1/2),
\end{eqnarray}
where $\zeta_{H}(s-1/2)$ is the generalized zeta function associated to the Hamiltonian operator $H$ and it definited by
\begin{eqnarray}
\zeta_{H}(s-1/2) = \sqrt{2\pi\mu^{2}} \sum_{i}\biggl(\frac{h_{i}^{2}}{2\pi\mu^{2}}\biggr)^{1/2-s}.
\end{eqnarray}
It is well-known that $\zeta_{\it M}(s)$ is analytic at $s=0$ [8, 11, 12]. So, $\zeta_{\it M}(0)$ is finite. Therefore either $\zeta_{H}(s-1/2)$ is analytic at $s=0$ and $\zeta_{\it M}(0)$ vanish or $\zeta_{H}(s-1/2)$ is not analytic at $s=0$ and in this case $\zeta_{H}(s-1/2)$ must have the same structure of simple poles as the gamma function $\Gamma(s)$, so that $\zeta_{\it M}(s)$ is analytic at $s=0$. Then the generalized zeta function associated to the operator $H$ can be written as
\begin{eqnarray}
\zeta_{H}(s-1/2) = F(s) + D(s)\Gamma(s),
\end{eqnarray} 
where $F(s)$ and $D(s)$ are analytic function at $s=0$. Note that $s$ is the regulator used in the analytic regularization procedure of the sum over zero-point energy.

The eq.(17) is evident when we employ the Laurent series expansion
$$
\zeta_{H}(s-1/2) = \frac{a_{-1}}{s} + a_{0} + a_{1}s + a_{2}s^{2} + ...,
$$
\begin{eqnarray}
\zeta_{H}(s-1/2) = \frac{a_{-1}2\sqrt{\pi}}{\Gamma(s-1/2)}\Gamma(s) + a_{0} + a_{1}s + a_{2}s^{2} + ...,
\end{eqnarray}
where $D(s)$ and $F(s)$ are self-evident. Since $\zeta_{H}(s-1/2)$ only has a simple pole the expansion is univocaly determined.

Now one differentiates the eq.(15) and using eq.(17) we obtain
$$
\zeta^{\prime}_{\it M}(s) = \frac{1}{2}\frac{\hbar}{\Omega}\frac{\Gamma(s-1/2)}{2\sqrt{\pi}\Gamma(s)} 
\biggl[\psi(s-1/2)F(s) + \psi(s-1/2)D(s) \Gamma(s) - \psi(s)F(s)  +
$$
\begin{eqnarray}
- {\bf \psi(s)\Gamma(s)D(s)} + F^{\prime}(s) + 
D^{\prime}(s)\Gamma(s) + {\bf D(s)\psi(s)\Gamma(s)} \biggr].
\end{eqnarray}
Observe that the terms in boldface are divergents, and they cancel them-selves. Finally we find
\begin{eqnarray}
\zeta^{\prime}_{\it M}(0) = -\frac{1}{2}\frac{\hbar}{\Omega}[F(0) + \psi(-1/2)D(0) + D^{\prime}(0)].
\end{eqnarray}
The result above explicitly show how the divergences are cancelled in the evaluation of the effective potential using the eq.(6). So, it is clear that hidden in $\zeta^{\prime}_{\it M}(0)$ there is a procedure to obtain the finite part of eq.(11).

\section{Equivalence}
Since the eq.(6) furnishes a finite result, there is, indeed, some procedure by mean of which one obtain, unambigously, the finite part of the theory. This procedure was employed by Salam-Strathdee [13] and proved by Bollini-Giambiagi to be mathematical solution for the problem [14]. The procedure is very simple: if $\Psi(s)$ is a function with just a simple pole in, say, $s=0$, then its finite part is defined as 
\begin{eqnarray}
F^{R} = \lim_{s\rightarrow 0}\biggl[\frac{\partial}{\partial s}\biggl(s\Psi(s)\biggr)\biggl].
\end{eqnarray}
It is worth to note that the eq.(21) has provided the necessary analytic continuation in order to restore the original theory.

Now, we know the zeta function $\zeta_{M}(s)$ is analytic at $s=0$. So, we can write it as
\begin{eqnarray}
\zeta_{M}(s) = -sf(s),
\end{eqnarray}
where $f(s)$ has a simple pole at $s=0$, in order to eq.(21) be satisfied.

Further, the minimum of the effective potential is the vacuum energy density which can be identified to the zero-point energy (ZPE) [15]. The zero-point energy is defined as
\begin{eqnarray}
\epsilon = \frac{\hbar T}{2\Omega}\sum_{i}h_{i}.
\end{eqnarray}
where $h_{i}$ are the eigenvalues of the Hamiltonian operator which are given by 
\begin{eqnarray}
h_{i}^{2} = p^{2} + V_{cl}^{\prime\prime}(\bar{\phi}).
\end{eqnarray}
Thus 
\begin{eqnarray}
\epsilon = \frac{\hbar T}{2\Omega}\sum\int d^{3}p \sqrt{p^{2} + V_{cl}^{\prime\prime}(\bar{\phi})}.
\end{eqnarray}
Of course the above sum (or integral) is clearly divergent.

In order to regularize the above expression, we exploit the concept of 
analytic continuation [16].
The idea is to replace the original expression with a divergent sum 
(integral) by a well behaved analytic function of a complex parameter. 
Now the sum (integral) is convergent and it can be evaluated formally 
without ambiguity. Then, the resulting expression can be continued 
analyticly to a value of the complex parameter which restores the original theory. The original ultraviolet divergence reasserts as pole at this value of the complex parameter. Subtraction of this pole at the end yields a finite result.
  
So, we replace $\epsilon$ by its analytic continuation, given by
\begin{eqnarray}
\epsilon = \frac{\hbar T}{2\Omega}\zeta_{H}(s-1/2)
\end{eqnarray}

In a analytic regularization procedure, as we have already said, the divergence appears simply as simple pole  at the physical value of the regularization parameter, s. Therefore $\epsilon(s)$ is a meromorphic function with simple pole at $s=0$ and then we can write it as
\begin{eqnarray}
\epsilon(s) = \frac{b_{-1}}{s} + \sum_{n=0}^{\infty}b_{n}s^{n}.
\end{eqnarray}
In order to restore the original theory, free of the divergences, we apply the procedure of eq.(21) on eq.(27). We get the renormalized zero-point energy 
\begin{eqnarray}
\epsilon^{R} = b_{0}.
\end{eqnarray}
Since $\epsilon^{R}$ is the vacuum energy density, it follows from eq.(6) and (22) that 
\begin{eqnarray}
f(s) = \epsilon(s).
\end{eqnarray}
Eq.(29) can be easily proved using eq.(15) and the identity
\begin{eqnarray}
\frac{2\sqrt{\pi}\Gamma[s]}{\Gamma[s-1/2]} = -\frac{1}{s} ,
\end{eqnarray}
so as to get
\begin{eqnarray}
\zeta_{M}(s) = - s\epsilon(s).
\end{eqnarray}
Comparing eq.(22) and eq.(31) we find the result of the eq.(29).
Now, using the eq.(31) we can write
\begin{eqnarray}
\epsilon(s) = - \frac{1}{s}\zeta_{M}(s).
\end{eqnarray}
Performing a expansion in Laurent series of the $\epsilon(s)$
\begin{eqnarray}
\epsilon(s) = - \zeta_{M}(0)\frac{1}{s} - \zeta^{\prime}_{M}(0) - \zeta^{\prime\prime}_{M}(0)\frac{s}{2} - ...  ,
\end{eqnarray}
and using the eq.(21) we obtain
\begin{eqnarray}
\epsilon^{R} = - \zeta^{\prime}_{M}(0).
\end{eqnarray}

Therefore, when we use the zeta function method we are, implicitly, making an analytic regularization procedure whose finite part is obtained as
\begin{eqnarray}
\lim_{s\rightarrow 0}\biggl[\frac{\partial}{\partial s}\biggl(s\epsilon(s)\biggr)\biggl].
\end{eqnarray}

Of course, the multiplication $\epsilon(s)$ by $s$ cancels the pole avoiding the possible divergences. Thus we can assert that the zeta function method is completely equivalent to the analytic regularization procedure to obtain the vacuum energy (or zero-point energy). We are not asserting the equivalence between the effective potential and the zero-point energy here, since this is a stronger assumption, and it is already well-known [17 - 19].

\section{Conclusion}
\h The regularization procedure of the minimum of the effective potential using zeta function method provides us finite result without the necessity of subtraction of any pole or addition of infinite counter-terms.

This is owing to zeta function is regular at $s=0$. In this manner dubious proceedings used in order to obtain of a finite result for the minimum of the effective potential are avoided, since as we explicitly showed the divergences are implicitly cancelled by zeta function method.

Zeta function method for the evaluation of the effective potential is implicitly equivalent to a analytic regularization procedure to obtain zero-point energy, whose finite part is obtained by the eq.(21). Further, eq.(21) furnishes the necessary analytic continuation to restore the original theory in the case of the analytic regularization of the zero-point energy.\\
\\
{\bf Aknowledgments}

We would like to thank the National Agency for Research (CNPq) (Brazil) for parcial support of this work.

\end{document}